\documentstyle[a4,12pt]{article}

\newcommand{\nc}[2]{\newcommand{#1}{#2}}
\newcommand{\ncx}[3]{\newcommand{#1}[#2]{#3}}
\ncx{\pr}{1}{#1^{\prime}}
\nc{\nl}{\newline}
\nc{\np}{\newpage}
\nc{\nit}{\noindent}
\nc{\be}{\begin{equation}}
\nc{\ee}{\end{equation}}
\nc{\ba}{\begin{array}}
\nc{\ea}{\end{array}}
\nc{\dsp}{\displaystyle}
\nc{\bit}{\bibitem}
\nc{\ct}{\cite}
\ncx{\dd}{2}{\frac{\partial #1}{\partial #2}}
\nc{\pl}{\partial}
\nc{\dg}{\dagger}
\nc{\ag}{\alpha}
\nc{\bg}{\beta}
\nc{\gam}{\gamma}
\nc{\Gam}{\Gamma}
\nc{\bgm}{\bar{\gam}}
\nc{\del}{\delta}
\nc{\Del}{\Delta}
\nc{\eps}{\epsilon}
\nc{\ve}{\varepsilon}
\nc{\th}{\theta}
\nc{\vt}{\vartheta}
\nc{\kg}{\kappa}
\nc{\lb}{\lambda}
\nc{\ps}{\psi}
\nc{\Ps}{\Psi}
\nc{\sg}{\sigma}
\nc{\spr}{\pr{\sg}}
\nc{\Sg}{\Sigma}
\nc{\rg}{\rho}
\nc{\fg}{\phi}
\nc{\Fg}{\Phi}
\nc{\vf}{\varphi}
\nc{\og}{\omega}
\nc{\Og}{\Omega}
\nc{\Kq}{\mbox{$K(\vec{q},t|\pr{\vec{q}\,},\pr{t})$ }}
\nc{\Kp}{\mbox{$K(\vec{q},t|\pr{\vec{p}\,},\pr{t})$ }}
\nc{\vq}{\mbox{$\vec{q}$}}
\nc{\qp}{\mbox{$\pr{\vec{q}\,}$}}
\nc{\vp}{\mbox{$\vec{p}$}}
\nc{\va}{\mbox{$\vec{a}$}}
\nc{\vb}{\mbox{$\vec{b}$}}
\nc{\Ztwo}{\mbox{\sf Z}_{2}}
\nc{\Tr}{\mbox{Tr}}
\nc{\lh}{\left(}
\nc{\rh}{\right)}
\nc{\ld}{\left.}
\nc{\rd}{\right.}
\nc{\cB}{\mbox{$^{\ast}\Og$ }}
\nc{\nil}{\emptyset}
\nc{\bor}{\overline}
\nc{\cD}{{\cal D}}
\nc{\tQ}{\tilde{Q}}
\nc{\wig}{\wedge}

\begin{document}

\pagestyle{empty}

\begin{flushright}
NIKHEF-H/94-29
\end{flushright}
\vspace{3ex}

\begin{center}

{\LARGE {\bf Supersymmetry and the Geometry of }} \\
\vspace{3ex}

{\LARGE {\bf Taub-NUT}} \\
\vspace{5ex}

{\Large  J.W.\ van Holten} \\
\vspace{3ex}

{\Large  NIKHEF-H} \\
\vspace{3ex}

{\Large Amsterdam NL} \\
\vspace{5ex}

{\bf September 22, 1994} \\
\vspace{15ex}

{\small {\bf Abstract}}

\end{center}

\nit
{\small
The supersymmetric extension of Taub-NUT space admits 4 standard
supersymmetries plus several additional non-standard ones. The geometrical
origin of these symmetries is traced, and their algebraic structure is
clarified. The result has applications to fermion modes in gravitational
instantons as well as in long-range monopole dynamics.}

\np

\pagestyle{plain}
\pagenumbering{arabic}

The low-energy dynamics of spinning particles in a curved space with metric
$g_{\mu\nu}(x)$ is described by the $d=1$ supersymmetric $\sg$-model
\ct{BM}-\ct{JW}

\be
L\, =\, \frac{1}{2} g_{\mu\nu}(x) \dot{x}^{\mu} \dot{x}^{\nu}\, +\,
        \frac{i}{2} \eta_{ab} \ps^a \frac{D\ps^b}{D\tau},
\label{1.1}
\ee

\nit
where $\eta_{ab}$ is the flat metric of tangent space: $\eta_{ab} = \del_{ab}$
for a Euclidean space, and diag$(-,+,...,+)$ for a Lorentzian space-time.
The Grassmann-variables $\ps^a$ transform as a tangent space vector and
describe the spin of the particle. More precisely, the anti-symmetric tensor
$S^{ab} = -i \ps^a \ps^b$ generates the internal part of the local
tangent-space rotations; also, for charged particles its components are
proportional to the tensor of electro-magnetic dipole moments \ct{JW2}. The
classical equations of motion of the theory can be cast in the form

\be
\frac{D^2 x^{\mu}}{D\tau^2}\, =\, \frac{1}{2}\, S^{ab}
          R_{ab\:\:\nu}^{\:\:\:\:\:\mu} \dot{x}^{\nu},  \hspace{3em}
\frac{DS^{ab}}{D\tau}\, =\, 0.
\label{1.2}
\ee

\nit
All derivatives here are covariant with respect to general co-ordinate
transformations and local tangent-space rotations. Equations (\ref{1.2}) also
hold in the quantum theory when interpreted as functional averages, and
therefore have a classical meaning in the sense of Ehrenfest's theorem.

An interesting consequence of these equations is, that particle spin can be
used as a probe of the geometry of space-time. Such information obtained
from particle spin is complementary to that from geodesics defined by the
worldlines of spinless test particles.

The constants of motion of a scalar particle in a curved space are determined
by the symmetries of the manifold, and are expressible in terms of the Killing
vectors and tensors. For a spinning particle a similar result holds, with
two modifications \ct{RJW1,GRvH}: first, the constants of motion related to a
given Killing vector generally contain spin-dependend parts; second, there are
Grassmann-odd constants of motion, related to supersymmetries, which do not
have a counterpart in the scalar model, although they do have interesting
geometrical origins. This is one of the ways in which spin can provide
additional information about geometry.

In this letter we study spinning particles in Taub-NUT space, which is a D=4
self-dual Euclidean space. In a special choice of co-ordinates the metric
takes the form

\be
ds^2\, =\, \lh 1 + \frac{2m}{r} \rh \lh dr^2 + r^2 d\th^2 + r^2 \sin^2 \th
           d\vf^2 \rh\, +\, \frac{4m^2}{\lh 1 + \frac{2m}{r}\rh}\,
           \lh d\ps + \cos \th d\vf \rh^2.
\label{1.2.1}
\ee

\nit
The physical applications of the Lagrangian (\ref{1.1}) with this metric are
either to fermions in a gravitational instanton background for $m > 0$ \ct{SH},
or to the long-range dynamics of interacting magnetic monopoles and their
fermion modes \ct{M,GM} for $m< 0$ and $r > 2|m|$. However, from the analysis
presented below it is clear that the supersymmetric extension also considerably
clarifies the structure of the purely bosonic theory. As such the method and
ideas seem to be of more general relevance. Some results on spinning Taub-NUT
space of which we make use were obtained in refs.\ct{V1,DB,V2}.

For the purpose of describing symmetries and conservation laws, the covariant
Hamiltonian formalism introduced in \ct{GRvH,JW3} is most useful. In this
formalism the basic phase-space variables are $(x^{\mu}, \Pi_{\mu}, \ps^a)$,
with $\Pi_{\mu}$ the covariant rather than the canonical momentum:

\be
\Pi_{\mu}\, =\, p_{\mu}\, -\, \frac{1}{2}\, \og_{\mu ab} S^{ab}.
\label{1.2.2}
\ee

\nit
For spinless scalar particles, the two are identical. In terms of these
variables the Hamiltonian takes the simple form

\be
H\, =\, \frac{1}{2}\, g^{\mu\nu} \Pi_{\mu} \Pi_{\nu}.
\label{1.3}
\ee

\nit
In the covariant phase space formulation the Poisson-Dirac bracket of two
scalar functions $A, B$ is given by

\be
\left\{ A, B \right\}\, =\, \cD_{\mu}A \dd{B}{\Pi_{\mu}}\, -\,\dd{A}{\Pi_{\mu}}
\cD_{\mu} B\, +\, R_{\mu\nu} \dd{A}{\Pi_{\mu}} \dd{B}{\Pi_{\nu}}\, +\,
i (-1)^{a_A}\, \dd{A}{\ps^a} \dd{B}{\ps_a},
\label{1.4}
\ee

\nit
with $R_{\mu\nu} = 1/2\, S_{ab} R^{ab}_{\mu\nu}$ the spin-valued Riemann
tensor, and

\be
\cD_{\mu} A\, =\, \pl_{\mu} A\, +\, \Gam_{\mu\nu}^{\:\:\:\:\:\lb} \Pi_{\lb}
        \dd{A}{\Pi_{\nu}}\, +\, \og^{\:\:a}_{\mu \:\:b} \ps^b \dd{A}{\ps^a}.
\label{1.5}
\ee

Before applying the formalism to the model (\ref{1.2.1}), we first observe that
the elementary constants of motion --those linear in the momentum-- for a
scalar particle in Taub-NUT space consist of the 3-$d$ total angular momentum
$\vec{J}$ and a quantity which, for negative mass models, can be interpreted as
the relative electric charge $q \equiv J_0$:

\be
J_A \, =\, R_A \cdot \Pi,
\label{1.5.1}
\ee

\nit
where the $R_A^{\mu}$ with $A = (0,...,3)$ and $\mu = (r,\th,\vf,\ps)$ are the
components of the four corresponding Killing vectors:

\be
\ba{lll}
R_0 & = & \lh 0, 0, 0, 1 \rh, \\
 & & \\
R_1 & = & \lh 0, - \sin \vf, - \cot \th\, \cos \vf, \csc \th\, \cos \vf \rh, \\
 & & \\
R_2 & = & \lh 0, \cos \vf, - \cot \th\, \sin \vf, \csc \th\, \sin \vf \rh, \\
 & & \\
R_3 & = & \lh 0, 0, 1, 0 \rh. \\
\ea
\label{1.5.2}
\ee

\nit
Expressions for the extension of these constants of motion to spinning
particles have been presented in \ct{V1,DB}. A fast method to obtain them is by
the following theorem:
\vspace{3ex}

\nit
{\em If on a manifold with metric $g_{\mu\nu}$ there exists a Killing vector
$R^{\mu}$, then the motion of a scalar particle on the manifold conserves the
quantity

\be
J\, =\, R^{\mu} \Pi_{\mu},
\label{1.5.3}
\ee

\nit
and the motion of a spinning particle with Lagrangian (1.1) conserves the
spin-dependend extension

\be
{\cal J}\, =\, R^{\mu} \Pi_{\mu}\, +\, \frac{1}{2}\, B_{ab}\, S^{ab},
\label{1.5.4}
\ee

\nit
where the base-space components of the anti-symmetric tensor $B_{\mu\nu}$ are
given in terms of the Killing vector by}

\be
B_{\mu\nu}\, =\, \frac{1}{2}\, \lh R_{\nu;\mu} - R_{\mu;\nu} \rh.
\label{1.5.5}
\ee

\nit
All constants of motion of this form are superinvariant:

\be
\left\{ {\cal J}, Q \right\}\, =\, 0,
\label{1.5.5.0}
\ee

\nit
and the Lie algebra defined by the Killing vectors:

\be
 R^{\;\mu}_{A\,;\nu} R_B^{\;\nu}\, -\, R^{\;\mu}_{B\,;\nu} R_A^{\;\nu}\, =\,
 f_{AB}^{\:\:\:\:\:C}\, R_C^{\;\mu},
\label{1.5.5.1}
\ee

\nit
is realized by the constants of motion ${\cal J}_A$ through the Poisson-Dirac
brackets:

\be
\left\{ {\cal J}_A, {\cal J}_B \right\}\, =\, f_{AB}^{\:\:\:\:\:C}\,
{\cal J}_C.
\label{1.5.5.2}
\ee

\nit
Note however, that in principle it is possible to add to the solution
$B_{\mu\nu}$ in eq.(\ref{1.5.5}) an improvement term $\bg_{\mu\nu}$, provided
it is covariantly constant:

\be
\bg_{\mu\nu ; \lb} = 0.
\label{1.5.5.3}
\ee

\nit
If such a tensor exists, the quantity

\be
\bg\, =\, \frac{1}{2}\, \bg_{ab} S^{ab},
\label{1.5.6}
\ee

\nit
is a constant of motion by itself; as a result there is some arbitrariness
in the definition of ${\cal J}$, eq.(\ref{1.5.4}). In general however,
improvement terms will change the algebraic properties (\ref{1.5.5.0}) and
(\ref{1.5.5.2}). Only specific choices of the improvement terms preserve the
algebraic structure. The Taub-NUT geometry provides an explicit example of this
phenomenon. In any case, without constraints on the algebra of Poisson-Dirac
brackets the solutions of the generalized Killing equations,
such as those found in \ct{V1,DB}, are not unique.

In addition to angular momentum and relative charge, the Taub-NUT geometry
admits a constant of motion known as the Runge-Lenz vector \ct{GM,GR}, which is
constructed out of a second-rank Killing tensor \ct{DR}. In the following we
show that this constant of motion is closely related to a number of
supersymmetries of the spinning Taub-NUT model.

The starting point of our analysis is the observation \ct{GRvH} that the
spinning particle action defined by (\ref{1.1}) or (\ref{1.3}) has a conserved
supercharge

\be
Q_f\, =\, f^{\mu}_{\:\:a}\, \Pi_{\mu} \ps^a\, +\, \frac{i}{3!}\, c_{abc}\,
          \ps^a \ps^b \ps^c,
\label{1.6}
\ee

\nit
provided the tensors $(f^{\mu}_{\:\:a}, c_{abc})$ satisfy the differential
constraints

\be
\ba{c}
D_{\mu} f^{\:\:a}_{\nu}\, +\, D_{\nu} f^{\:\:a}_{\mu}\, =\, 0, \\
  \\
D_{\mu} c_{abc}\, +\, R_{\mu\nu ab} f^{\nu}_{\:\:c}\, +\, R_{\mu\nu bc}
        f^{\nu}_{\:\:a}\, +\, R_{\mu\nu ca} f^{\nu}_{\:\:b}\, =\, 0.
\ea
\label{1.7}
\ee

\nit
The corresponding supersymmetry transformations of the bosonic co-ordinates are

\be
\del x^{\mu}\, =\, - i \eps\, f^{\mu}_{\:\:a}(x) \ps^a.
\label{1.8}
\ee

\nit
A particularly simple solution is provided by the vierbein field:
$f^{\mu}_{\:\:a} = e^{\mu}_{\:\:a}$, $c_{abc} = 0$. In eq.(\ref{1.6}) this
gives the standard supercharge $Q = \Pi \cdot \ps$, confirming that the
theories defined by (\ref{1.1}) always possess at least one supersymmetry.

However, depending on the geometry of the manifold, other solutions can exist.
In particular, if the metric admits a tensor $f_{\mu\nu} = f_{\mu a}
e_{\nu}^{\:\:a}$  of Killing-Yano type, it implies the presence of a new
supersymmetry which anti-commutes with $Q$ \ct{GRvH}. A tensor $f_{\mu\nu}$ is
called Killing-Yano if it is anti-symmetric and its field strength satifies

\be
H_{\mu\nu\lb}\, \equiv\, \frac{1}{3}\, \lh f_{\mu\nu ;\lb} + f_{\nu\lb ;\mu}
                            + f_{\lb\mu ;\nu} \rh\, =\, f_{\mu\nu;\lb},
\label{1.9}
\ee

\nit
which follows in fact from the first eq.(\ref{1.7}). Differentiation of
$H_{\mu\nu\lb}$ and use of the Ricci identity then shows, that $ -2H_{abc}$ is
a solution of the second equation (\ref{1.7}) for $c_{abc}$. Hence the
existence of a Killing-Yano tensor of the bosonic manifold is equivalent to
the existence of a supersymmetry for the spinning particle with supercharge

\be
Q_f\, =\, f^{\mu}_{\:\:a}\, \Pi_{\mu} \ps^a\, -\, \frac{i}{3}\, H_{abc}\,
          \ps^a \ps^b \ps^c, \hspace{3em}
\left\{ Q, Q_f \right\}\, =\, 0.
\label{1.10}
\ee

\nit
In the Taub-NUT geometry (\ref{1.2.1}) four Killing-Yano tensors are known to
exist \ct{GR}. Three of these, denoted by $f_{i}$, $i = (1,2,3)$, are
special because they are covariantly constant; therefore they satisfy
eq.(\ref{1.9}) trivially by having vanishing field strength. In 2-form
notation and in the co-ordinates (\ref{1.2.1}) the explicit expressions for
the $f_i$ are

\be
f_i\, =\, 4 m \lh d\ps + \cos \th d\vf \rh \wedge dx_i\, -\,
          \ve_{ijk}\, \lh 1 + \frac{2m}{r} \rh dx_j \wedge dx_k ,
\label{1.11}
\ee

\nit
where the $dx_i$ are standard expressions in terms of the 3-dimensional
spherical co-ordinates $(r,\th,\vf)$. The corresponding supercharges have the
simple form

\be
Q_i\, =\, f_{i\;\;a}^{\;\mu} \Pi_{\mu} \ps^a,
\label{1.12}
\ee

\nit
and together with $Q_0 = Q$ they realize the $N = 4$ supersymmetry algebra

\be
\left\{ Q_A, Q_B \right\}\,=\, -2i \del_{AB}\, H, \hspace{2em}
       (A,B)\, =\, 0,...,3.
\label{1.13}
\ee

\nit
Indeed, the $f_i$ define three anti-commuting complex structures on the
Taub-NUT manifold, their components realizing the quaternion algebra

\be
f_i f_j\, +\, f_j f_i\, =\, -2 \del_{ij}, \hspace{3em}
f_i f_j\, -\, f_j f_i\, =\, 2 \ve_{ijk} f_k.
\label{1.14}
\ee

\nit
Clearly, the existence of the Killing-Yano tensors is linked to the
hyper-K\"{a}hler geometry of the manifold, and our argument shows directly
the relation between this geometry and the $N = 4$ supersymmetry of the theory.
We can also give a {\em physical} interpretation of these particular
Killing-Yano tensors: they correspond to components of the spin which are
separately conserved. Namely, three constants of motion are obtained by
rewriting the 2-forms (\ref{1.11}) in terms of the spin co-ordinates:

\be
S_i\, =\, \frac{i}{4}\, f_{i\,ab} \ps^a \ps^b\, =\,
          -\frac{1}{4}\, f_{i\,ab} S^{ab}, \hspace{3em}
\left\{ S_i, H \right\}\, =\, 0,
\label{1.15}
\ee

\nit
which realize an $SO(3)$ Lie-algebra:

\be
\left\{ S_i, S_j \right\}\, =\, \ve_{ijk}\, S_k.
\label{1.15.1}
\ee

\nit
These constant 2-forms transform as a vector under rotations generated by the
total angular momentum ${\cal J}_i$. Obviously, they provide improvement
terms of the type (\ref{1.5.6}) for the components of total angular momentum.
It follows, that we can define an improved form of the angular momentum

\be
{\cal I}_i\, =\, {\cal J}_i\, -\, S_i,
\label{1.15.2}
\ee

\nit
with the property that it preserves the algebra:

\be
\left\{ {\cal I}_i, {\cal I}_j \right\}\, =\, \ve_{ijk}\, {\cal I}_k,
\label{1.15.3}
\ee

\nit
and that it {\em commutes} with the $SO(3)$ algebra generated by the spins
$S_i$. Equivalently, we can combine these two $SO(3)$ algebras to obtain the
generators of a conserved $SO(4)$ symmetry among the constants of motion,
a standard basis for which is spanned by $M^{\pm}_i = {\cal I}_i \pm S_i $.
Under this $SO(4)$ the supercharges $Q_A$ transform as a 4-vector:

\be
\ba{ll}
\left\{ M^+_i, Q \right\}\, =\, 0, &
                                  \left\{ M^-_i, Q \right\}\, =\, Q_i, \\
 & \\
\left\{ M^+_i, Q_j \right\}\, =\, \ve_{ijk}\, Q_k, &
                        \left\{ M^-_i, Q_j \right\}\, =\, -\del_{ij} Q, \\
\ea
\label{1.15.4}
\ee

\nit
Obviously these $SO(4)$ transformations preserve the $N = 4$ supersymmetry
algebra (\ref{1.13}). Note also, that the second equation above implies
that the supercharges $Q_i$ are $Q$-exact and hence trivially superinvariant.
We conclude that the algebra of constants of motion $(M^{\pm}_i, Q_A, H)$
closes. It represents the universal part of the algebra of any model
with $N = 4$ supersymmetry.

However, in the case of the Taub-NUT model, in addition to the four standard
supersymmetries whose physical and geometrical origin has been clarified, the
metric is known to possess a {\em fourth} Killing-Yano tensor, which is not
trivial and leads to new constants of motion. Its components are contained in
the 2-form

\be
Y\, =\, 4 m \lh d\ps + \cos \th d\vf \rh \wedge dr\, +\, 4r \lh r + m \rh
     \lh 1 + \frac{r}{2m} \rh\, \sin \th\, d\th \wedge d\vf .
\label{1.18}
\ee

\nit
The field strength has one independend non-vanishing component, given by

\be
H_{r\th\vf}\, =\, 2 \lh 1 + \frac{r}{2m} \rh\, r \sin \th .
\label{1.19}
\ee

\nit
The supercharge $Q_Y$ constructed from $Y$ is a scalar under rotations, but
it is not invariant under the $SO(3)$ transformations generated by the $S_i$.
Instead, by computing the brackets we identify another triplet of conserved
supercharges

\be
\Og_i\, =\, \left\{ M^-_i,  Q_Y \right\}\,
        =\, -2\,\left\{ S_i,  Q_Y \right\}.
\label{1.20}
\ee

\nit
Although these supercharges provide solutions to eqs.(\ref{1.6}), (\ref{1.7}),
they are  not superinvariant, and therefore they do not imply
the existence of any new Killing-Yano tensors. Of course, $Q_Y$ itself is
superinvariant by construction; but its brackets with the other supercharges
$Q_i$ do not vanish: they generate a triplet of new constants of motion

\be
K_i\, =\, \left\{ Q_Y, Q_i \right\}.
\label{1.21}
\ee

\nit
The explicit expressions for the components of $K_i$ are constructed from
(\ref{1.11}), (\ref{1.18}) and (\ref{1.19}) according to the formulas given in
\ct{GRvH,JW3}. The results can be summarized briefly as follows. \nl
{\em a.} The Killing tensors $K_{i\, \mu\nu}$ on the left-hand side of
(\ref{1.20}) are those first obtained in \ct{DR} and identified with a
conserved vector of Runge-Lenz type for the monopole scattering problem in
\ct{GR}. \nl
{\em b.} There is a contribution to $K_i$ linear in the spin $S^{ab}$; however,
there is no contribution quadratic in the spin, as observed in \ct{V2}.
Comparing with the general expressions given in \ct{GRvH} this can be seen to
follow from the fact that the complex structures $f_i$ are covariantly constant
and the self-duality of the Taub-NUT geometry, which together imply the
identities

\be
\eps^{abcd}\, R_{cd \mu\nu}\,  f^{\mu}_{i\:a}\,f^{\nu}_{j\:b}\, =\, 0,
\hspace{3em}
c_{i\,abc}\, =\, 0.
\label{1.22}
\ee

\nit
The Jacobi identities now imply, that the supercharges $\Og_i$ are the lowest
(fermionic) components of a superdoublet, the highest (bosonic) components of
which form the Runge-Lenz vector, which is generated from them by a
supersymmetry transformation:

\be
K_i\, =\, - \left\{ Q, \Og_i \right\}.
\label{1.23}
\ee

\nit
The remaining brackets between supercharges give constants of motion which are
polynomials in the ones already found. In particular, the bracket of $Q_Y$ with
itself gives

\be
\left\{ Q_Y, Q_Y \right\}\, =\, 2i H\, +\, i \lh \vec{{\cal J}}^{\,2} -
        {\cal J}_0^2 \rh,
\label{1.24}
\ee

\nit
where $\vec{{\cal J}}$ and ${\cal J}_0$ are the total angular momentum and the
relative electric charge of the monopoles (for $m < 0$) introduced before,
eqs.(\ref{1.5.2}) and (\ref{1.5.4}).
\vspace{2ex}

Finally we briefly discuss the quantum mechanics of the spinning particle in
Taub-NUT space. Using the Lagrangian (1.1) in a path-integral, the symmetries
investigated in this paper lead directly to relations between matrix elements
in the co-ordinate representation. We can also investigate these relations at
the level of operators by using the correspondence principle

\be
\ps^a\, \rightarrow\, i\sqrt{ \frac{\hbar}{2} }\, \gam_5 \gam^a, \hspace{3em}
\Pi_{\mu}\, \rightarrow\, -i \hbar D_{\mu}\, =\, -i \hbar \lh \pl_{\mu} -
                          \frac{1}{2} \og_{\mu ab} \sg^{ab} \rh.
\label{1.25}
\ee

\nit
Then the supercharges correspond to generalizations of the Dirac operator:

\be
Q_f\, \rightarrow\, i \sqrt{\frac{\hbar}{2}}\, \gam_5 D_f\, \equiv\,
      \frac{\hbar^{3/2}}{\sqrt{2}}\, \gam_5 \gam^a\, \lh f^{\mu}_{\:\:a}
      D_{\mu} - \frac{1}{3!} c_{abc} \sg^{bc} \rh,
\label{1.26}
\ee

\nit
whilst their anti-commutators, representing the operators for $H$ and $K_i$,
are given by the corresponding Laplacians. The $N = 4$ supersymmetry now
implies the existence of 4 Dirac operators $D_A$ with the property

\be
\left\{ D_A, D_B \right\}\, = - 2 \hbar^2 \del_{AB} \, \Box^{cov}.
\label{1.27}
\ee

\nit
At least two operators from the set $(D_A, \gam_5 D_A)$ can always be
diagonalized simultaneously. However, a zero-mode of any of these Dirac
operators is a zero-mode of the covariant Laplacian and vice-versa, therefore
the kernels of all four operators $D_A$ coincide.

If a spinor transforms non-trivially under the (global) $SO(4)$, more
possibilities arise. For example, a four-vector of spinors $\Ps_A$ might
satisfy one or both of the $SO(4)$-invariant conditions

\be
D \cdot \Ps\, =\, 0,  \hspace{3em} D_A \Ps_B\, -\, D_B \Ps_A\, =\, 0.
\label{1.28}
\ee

\nit
If both conditions are satisfied, the $\Ps_A$ are zero-modes of the Laplacian
and hence of all Dirac operators $D_A$ simultaneously. However, if the $\Ps_A$
are non-zero modes of the Laplacian (finite mass particles), then they realize
a partially broken $N = 4$ supersymmetry. Clearly the structure of the space of
solutions may become quite intricate.

Finally, the zero-modes of the fifth Dirac-type operator $D_Y$ coincide with
those of the other Dirac operators if and only if they are extremal in the
sense that the square of their relative charge equals the total angular
momentum:

\be
\vec{{\cal J}}^{\, 2}\, \Ps\, =\, {\cal J}_0^2\, \Ps,
\label{1.29}
\ee

\nit
or $j(j + 1)\, -\, q^2\, =\, 0$ , where $j$ is the quantum number of total
angular momentum, and $q$ that of the relative electric charge. If not, the
zero-modes of the Laplacian correspond to doublets of the finite eigenvalue
equations

\be
 D_Y\, \Ps\, =\, \lb\, \Phi, \hspace{3em}
 D_Y\, \Phi\, =\, - \lb\, \Ps,
\label{1.30}
\ee

\nit
for $j(j + 1)\, -\, q^2\, >\, 0$. For $j(j + 1)\, -\, q^2\, <\, 0$ there are no
normalizable solutions of eqs.(\ref{1.30}) with real $\lb$. These bounds
correspond to those found to distinguish between monopole bound states and
scattering states in \ct{GM}.

\vspace{4ex}

\nit
{\bf Acknowledgement}\nl

\nit
The research described in this paper is supported in part by the Human
Capital and Mobility Program through the network on {\em Constrained
Dynamical Systems}.

\end{document}